\documentclass[aps,pre,twocolumn,showpacs]{revtex4}

\newcommand{\beq}{\begin{equation}}
\newcommand{\eeq}{\end{equation}}
\newcommand{\bea}{\begin{eqnarray}}
\newcommand{\eea}{\end{eqnarray}}

\begin{document}

\title{Eulerian spectral closures for isotropic turbulence using a time-ordered 
fluctuation-dissipation relation} 

\author{W. D. McComb}
\email{w.d.mccomb@ed.ac.uk}
\author{K. Kiyani}
\affiliation{School of Physics, University of Edinburgh, Mayfield Road,
Edinburgh EH9 3JZ, UK}

\date{\today}

\begin{abstract}
Procedures for time-ordering the covariance function, as given in a
previous paper (K. Kiyani and W.D. McComb \emph{Phys. Rev. E}
\textbf{70}, 066303 (2004)), are extended and used to show that the
response function associated at second order with the Kraichnan-Wyld
perturbation series can be determined by a local (in wavenumber) energy
balance. These time-ordering procedures also allow the two-time formulation to be
reduced to time-independent form by means of exponential approximations
and it is verified that the response equation does not have an infra-red
divergence at infinite Reynolds number. Lastly, single-time
Markovianised closure equations (stated in the previous paper \cite{Kiyani04})
are derived and shown to be compatible with the Kolmogorov distribution
without the need to introduce an \emph{ad hoc} constant.
\end{abstract}


\maketitle

\section{Introduction}

In a previous paper \cite{Kiyani04}, the Kraichnan-Wyld perturbation
expansion \cite{Kraichnan59b,Wyld61} was used to justify the
introduction of a renormalized response function connecting two-point 
covariances at different times. The resulting relationship was
specialized by suitable choice of initial conditions to the form of a
fluctuation-dissipation relation (FDR). This was further developed to
reconcile the time-symmetry of the covariance with the causality of the
response function by the introduction of time-ordering along with a
counter-term. We pointed out that this formulation provides a solution
to an old problem in turbulence theory: that of representing the time-dependence of the
covariance and response by exponential forms \cite{Kraichnan64c,Leslie73}. 
We showed that the derivative (with respect to difference time) of the covariance with
this time-ordering now vanishes at the origin.  This allows one to study the relationships
between two-time spectral closures and time-independent theories such as
the Fokker-Planck theory of Edwards \cite{Edwards64} or the
more recent renormalization group approaches. We also showed that the
renormalized response function is transitive with respect to
intermediate times and reported a new Langevin-type equation for
turbulence.

In this paper we interpret the second-order response function as a
mean-field propagator and show that in addition to propagating two-time
covariances it also links single-time covariances. We then make use of
its newly established properties to re-derive the local-energy transfer
(LET) response equation \cite{McComb92} and show that it now contains a counter term
which removes the singularity of previous propagator equations at
$t=t'$. We also introduce a partial-propagator representation and
hence reformulate the LET statistical equations. Furthermore we specialize the
two-time equations to time-independent form by introducing exponential time
dependences and show that the
closure is well-behaved in the limit of infinite Reynolds number.
Lastly, by Markovianizing time-history 
integrals, we end up with a Langevin-type theory which is compatible
with the Kolmogorov spectrum without the need to introduce 
\emph{ad hoc} constants as in the case of the EDQNM model \cite{Orszag70} for example.

We begin by reviewing the subject of turbulence closures and then go on
to consider various aspects of applying the FDR to nonequilibrium,
macroscopic problems such as fluid turbulence. We begin by stating the basic
equations. 

\subsection{The basic equations}

Following standard practice in this topic \cite{McComb90a}, we consider the 
solenoidal Navier-Stokes
equation (NSE) in wavenumber ($k$) space, as follows:
\begin{equation}
	\left( \frac{\partial}{\partial t} + \nu_0 k^2 \right)
	u_\alpha (\mathbf{k},t) = M_{\alpha\beta\gamma} (\mathbf{k})
	\int \mbox{d}^3j \, u_\beta (\mathbf{j},t) u_\gamma
	(\mathbf{k}-\mathbf{j},t), \label{nse}
\end{equation}
while the continuity equation for incompressible fluids is
\begin{equation}
	k_\alpha u_\alpha (\mathbf{k},t) = 0.
\end{equation}
The inertial transfer operator $M_{\alpha\beta\gamma}(\mathbf{k})$ is given by
\begin{equation}
	M_{\alpha\beta\gamma} (\mathbf{k}) = (2i)^{-1} [k_\beta
	P_{\alpha\gamma}(\mathbf{k}) + k_\gamma
	P_{\alpha\beta}(\mathbf{k})],
\end{equation}
while the projector $P_{\alpha\beta} (\mathbf{k})$ is expressed in
terms of the Kronecker delta as
\begin{equation}
	P_{\alpha\beta}(\mathbf{k}) = \delta_{\alpha\beta} -
	\frac{k_\alpha k_\beta}{|\mathbf{k}|^2}.
\label{project}
\end{equation}
In order to introduce a statistical treatment, we shall denote the
operation of performing an ensemble average by angle brackets, thus
$\langle \cdots \rangle$, and restrict our attention to isotropic, homogeneous
turbulence, with energy dissipation rate $\varepsilon$ and zero mean
velocity. As a result of this restriction, the covariance of the
fluctuating velocity field takes the form
\begin{equation}
	\langle u_\alpha (\mathbf{k},t) u_\beta (\mathbf{k'},t')
	\rangle = C(k;t,t') P_{\alpha\beta}(\mathbf{k}) \delta
	(\mathbf{k}+\mathbf{k'}),
\end{equation}
where $\alpha,\,\beta=1,2$ or $3$. The corresponding single-time quantity
may be written as
\beq
C(k;t,t)=C(k,t),
\eeq
where the single-time one-point covariance $C(k,t)$ may be interpreted as a spectral density and 
is related to the energy spectrum by
\begin{equation}
	E(k,t) = 4 \pi k^2 C(k,t). \label{eq}
\end{equation}
Using (\ref{nse}), we can also obtain an equation describing the
energy balance between spatial modes.  To do this, we first multiply each term in
(\ref{nse}) by $u_\sigma (-\mathbf{k},t)$.  Then we form a second
equation from (\ref{nse}) for $u_\sigma (-\mathbf{k},t)$, multiply
this by $u_\alpha(\mathbf{k},t)$, add the two resulting equations
together, integrate over $\mathbf{k'}$ and average the final expression.  This leaves us with 
\begin{eqnarray}
	\lefteqn{\left( \frac{\partial}{\partial t} + 2 \nu_0 k^2
	\right) P_{\alpha\sigma}(\mathbf{k}) C(k,t) } \nonumber \\
	& = & M_{\alpha\beta\gamma} (\mathbf{k}) \int \mbox{d}^3j
	C_{\sigma\beta\gamma} (\mathbf{-k, j, k-j};t)  \nonumber \\
	&& - M_{\sigma\beta\gamma} (\mathbf{k}) \int \mbox{d}^3j
	C_{\sigma\beta\gamma} (\mathbf{k},\mathbf{j},
	\mathbf{-k-j};t), \label{balance_deriv}
\end{eqnarray}
where 
\beq
C_{\alpha\beta\gamma}(\mathbf{k,j,-k-j};t) =
\langle u_\alpha (\mathbf{k},t) u_\beta
(\mathbf{j},t) u_\gamma (\mathbf{-k-j},t) \rangle,
\eeq
 and where we have also used the property
\beq
M_{\alpha\beta\gamma} (\mathbf{-k}) = -
M_{\alpha\beta\gamma}(\mathbf{k}).
\eeq
If we then take the trace of (\ref{balance_deriv}) by setting $\sigma = \alpha$ and
summing over $\alpha$ (noting that $ Tr\, P_{\alpha\beta} = 2$), and
multiply each term in (\ref{balance_deriv}) by $2\pi k^2$, we obtain
\begin{equation}
	\left( \frac{\partial}{\partial t} + \nu_0 k^2 \right) E(k,t)
	= T(k,t),
\end{equation}
where
\begin{eqnarray}
	T(k,t) & = & 2 \pi k^2 M_{\alpha\beta\gamma} (\mathbf{k}) \int
	\mbox{d}^3j \, \left\{ C_{\alpha\beta\gamma}
	(-\mathbf{k},\mathbf{j}\mathbf{k}-\mathbf{j},t) \right. \nonumber \\
	& & \left. - C_{\alpha\beta\gamma}
	(\mathbf{k},\mathbf{j},-\mathbf{k}-\mathbf{j},t) \right\}. 
\end{eqnarray}
Evidently, in order to solve for the energy spectrum (or, second-order
moment) we need to know the third-order moment. Hence we are faced with
a hierarchy of statistical equations to be solved; and this is, of
course, the notorious closure problem.
 
\subsection{Eulerian statistical closures for isotropic turbulence}

In order to study isotropic turbulence, we have to add a noise term or
stirring force to the right hand side of the NSE, as given by (\ref{nse}).
Denoting this term by $f_{\alpha}(\mathbf{k},t)$, we specify it in terms
of its distribution, which we take to be Gaussian, and its covariance,
which we take to be of the form
\beq
\langle f_{\alpha}(\mathbf{k},t)f_{\beta}(\mathbf{k'},t') \rangle =
W(k)(2\pi)^3 P_{\alpha\beta}(\mathbf{k})\delta(\mathbf{k}+\mathbf{k'})\delta(t+t').
\eeq
We note that $W(k)$ is a measure of the rate at which the stirring
forces do work on the fluid and for stationarity must satisfy the
condition
\beq
\int_0^{\infty}4\pi k^{2}W(k) dk = \varepsilon = \int^{\infty}_{0}2\nu
k^{2}E(k)dk.
\label{work}
\eeq
The perturbative treatment of the equations of motion is based on an
expansion about a Gaussian zero-order velocity obtained by solving the
NSE with the nonlinear term set to zero. The resulting expansion shows 
clearly \cite{Wyld61} the
effect of nonlinear mixing such that any correction to the zero-order
field must have a non-Gaussian distribution, which indeed is implied by
the existence of the third-order moment (and the existence of
inter-modal energy transfer). Renormalization of the perturbation
expansion corresponds to either partial summation or term-by-term
reversion: for details reference should be made to the paper by Wyld
\cite{Wyld61} and the books by McComb \cite{McComb90a} and 
Leslie \cite{Leslie73}. Our present interest is restricted to the second-order 
equation for the velocity covariance, which is obtained by this procedure, thus:
\begin{eqnarray}
\lefteqn{\left[\frac{\partial}{\partial t} + \nu k^{2}\right]C(k;t,t')}
\nonumber  \\
 & = & \int d^{3}j
L({\bf  k},{\bf  j})\left[\int^{t'}_{0}d s R(k;t',s)C(j;t,s)C(|{\bf  k}-{\bf 
j}|;t,s) \right.  \nonumber \\ 
& - & \left. \int^{t}_{0}d s R(j;t,s)C(k;s,t')C(|{\bf 
k}-{\bf  j}|;t,s)\right],
\label{dia}
\end{eqnarray}
and on the time diagonal
 \begin{eqnarray}
 & &  \!\!\!\!\!\!\!\!\!\!\!\!\!\!\! 
 \left(\frac{\partial}{\partial t}+2\nu k^{2}\right)C\left(k,t\right)
 =2\int d^{3}jL\left(\mathbf{k},\mathbf{j}\right) \nonumber \\
& & \times \int_{0}^{t}dsR\left(k;t,s\right)
 R\left(j;t,s\right)R\left(\left|\mathbf{k}-\mathbf{j}\right|;t,s\right)\nonumber \\
& &  \!\!\!\!\!\!\!\!\!\!\!\!\!\!\! 
\times \left[C\left(j,s\right)C\left(\left|\mathbf{k}-\mathbf{j}\right|,s\right)- 
C\left(k,s\right)C\left(\left|\mathbf{k}-\mathbf{j}\right|,s\right)\right],
 \label{diag}  
 \end{eqnarray}
where the coefficient $L({\bf  k}, {\bf  j})$ is given by:
\beq
L({\bf  k},{\bf  j}) =-2M_{\alpha\beta\gamma}(\mathbf{k})M_{\beta\alpha\delta}(\mathbf{j})
P_{\gamma\delta}(\mathbf{k-j}). 
\eeq
This may be evaluated in terms of the scalar magnitudes $k$, $j$ and
$\mu=\cos \theta$, where $\theta$ is the angle between the two
wavevectors $\mathbf{k}$ and $\mathbf{j}$, thus:
\beq
L({\bf  k},{\bf  j}) = \frac{\left[\mu(k^{2} + j^{2}) - k j(1 +
2\mu^{2})\right](\mu^{2}-1)k j}{k^{2} + j^{2} - 2k j\mu}.
\eeq
It should also be noted that the coefficient $L({\bf  k},{\bf  j})$ is
symmetric under interchange of the two wavevectors: we shall use this fact
presently to demonstrate conservation of energy.

At this stage we should note that for this to be a closed set of
equations for the covariance $C$, one has to have an additional equation
to determine the response function $R$.
Equation (\ref{dia}) was originally derived by Kraichan.
This closure was completed by an equation for the response-function
$R(k; t, t')$, and is known as the \emph{direct interaction
approximation} or DIA. 
The basic \emph{ansatz} of DIA is that there exists a response function such
that 
\beq
\delta u_{\alpha}({\bf  k}, t) =
\int^{t}_{-\infty}\hat{R}_{\alpha\beta} ({\bf  k}; t, t')\delta
f_{\beta}({\bf  k}, t')dt',
\label{respdefine}
\eeq
and that this {\em infinitesimal response} function can be renormalized.
The resulting response equation is
\begin{eqnarray}
& & \left[\frac{\partial}{\partial t} + \nu k^{2}\right]R(k;t,t') 
+ \int d^{3}{j} L({\bf  k},{\bf  j})  \nonumber \\
& & \times \int^{t}_{t'}dt'' R(k;t'',t')
R(j;t,t'')C(|{\bf  k}-{\bf 
j}|;t,t'') \nonumber \\ 
& & = \delta(t-t').
\label{rdia}
\end{eqnarray} 
Later Edwards derived a time-independent covariance equation by the
self-consistent introduction of a generalized Fokker-Planck equation as
an approximation to the (rigorous) Liouville equation. We shall refer to
this theory as EFP, and this along with the more general
\emph{self-consistent field} (SCF) theory of Herring \cite{Herring65}
and the DIA make up our trio of pioneering spectral closures. Further
discussion can be found in the books \cite{Leslie73},\cite{McComb90a}.
In the literature, much attention has been given to the fact that, although these
theories have many satisfactory features, they are all incompatible with
the Kolmogorov (K41) power law for the energy spectrum $E(k)$
\cite{Kolmogorov41a}. However, in
the present paper we shall concentrate on only a few key points.
The first of these is that the covariance equation of the EFP
theory can be shown to be equivalent, for the stationary case, to the
second-order truncation of the Kraichnan-Wyld perturbation theory, if we
assume exponential time dependences. That is, the EFP covariance
equation can be obtained by substituting into
equation (\ref{dia}) the following assumed time dependences:
\beq
C(k,t-t') = C(k)\exp\{-\omega(k)|t-t'|\};
\eeq
and
\begin{eqnarray}
R(k,t-t') & = & \exp\{-\omega(k)(t-t')\} \quad \mbox{for} \quad t \geq t'; \nonumber \\
& = & 0 \hspace{3.1cm} \mbox{for} \quad t<t'.
\end{eqnarray}
Then, integrating the right hand side of equation (\ref{dia}) over intermediate times, one obtains
(with some re-arrangement):
\beq
W(k)-2\nu k^{2}C(k)= \int d^{3}j L(k,j)
\frac{C(|\mathbf{k}-\mathbf{j}|)\left[C(k)-C(j)\right]}
{\omega(k)+\omega(j)+\omega(\mathbf{k}-\mathbf{j}|)},
\label{efp}
\eeq
where we have added the term $W(k)$ to the energy balance in order to
sustain the turbulence against viscous dissipation. Equation (\ref{efp}) is just the
form originally derived by Edwards \cite{Edwards64}.

This simple form is helpful in understanding certain properties, such as
the  conservation of energy by the nonlinear term and the behaviour of
the system in the limit of infinite Reynolds number. For instance,
integrating both sides of equation (\ref{efp}) with respect to
$\mathbf{k}$ and invoking equation (\ref{work}) leads to:
\beq
\varepsilon - \varepsilon = 0,
\eeq
where the vanishing of the right hand side results from the antisymmetry
of the integrand under interchange of $k$ and $j$. This result helps us
to interpret the EFP response function or eddy decay rate $R(k)$
which takes the form \cite{Edwards64},
\beq
R(k)= \int d^{3}j L(k,j)
\frac{C(|\mathbf{k}-\mathbf{j}|)}
{\omega(k)+\omega(j)+\omega(\mathbf{k}-\mathbf{j}|)}.
\eeq
At the time this was interpreted as allowing one to write the energy
balance equation as:
\bea
& & W(k)-2\nu k^{2}C(k) = R(k)C(k) \nonumber \\
& & - \int d^{3}j L(k,j)
\frac{C(|\mathbf{k}-\mathbf{j}|)C(j)}
{\omega(k)+\omega(j)+\omega(\mathbf{k}-\mathbf{j}|)},
\eea
or, the eddy decay rate represented the loss of energy from mode $k$ due
to energy transfers to all other modes. The situation is more
complicated for DIA, but the analogous comment has been made by
Kraichnan \cite{Kraichnan59b} that the energy loss from mode $k$ is
directly proportional to the excitation of that mode, viz., $C(k;t,t')$.

Later, it was pointed out that an \emph{ad hoc} modification could be
made to EFP by noting that the entire energy transfer term (i.e. the
right hand side of (\ref{efp})) acts as an energy loss in some regions of
wavenumber space whereas in others it behaves as an input. This led to a
definition of the response which was compatible with the Kolmogorov
spectrum \cite{McComb74a,McComb76a} and this was subsequently generalised
to the two-time \emph{local energy transfer} or LET theory
\cite{McComb78}.

\subsection{Fluctuation-dissipation relations (FDR)}

It is well known that the response of microscopic systems in thermal
equilibrium to small perturbations is fully determined by the
covariance of the system fluctuations about equilibrium. In our
present notation, the relationship may be written as
\beq
C(k; t,t') = R(k; t't')C(k;t',t'),
\label{fdr}
\eeq
which is the \emph{fluctuation-dissipation relation} or FDR. This
result was extended by Kraichnan to nonlinear dynamical systems in
thermal equilibrium \cite{Kraichnan58, Kraichnan59a} and by Leith
\cite{Leith75} to inviscid two-dimensional chaotic flow. Also, Deker and
Haake \cite{Deker75} give several examples of classical processes for
which a FDR will hold and these include (of particular relevance to the
present discussion) forced viscous flows where the stationary probability
distribution is Gaussian. In realistic cases, such flows will have a
non-Gaussian distribution due to nonlinear mode coupling. However, one
case of interest arises in a pioneering application of renormalization
group methods to stirred fluid motion \cite{Forster77}, where a
fluctuation-dissipation relationship is found to hold in the limit
$k\rightarrow 0$. That is, the long-wavelength behaviour at lowest
nontrivial order of perturbation theory.

The paper by Leith is particularly interesting. While recognising that
the FDR cannot apply exactly to real fluid turbulence, it puts forward
rather convincing heuristic arguments for believing that it could be a
reasonable approximation. Also, it cites the investigation of Herring
and Kraichnan \cite{Herring72} in support of this view. Here the
non-stationary generalization of the SCF \cite{Herring66}, which differs
only from DIA in the use of the FDR, gives very similar results to it.
We shall discuss this use of FDR in more detail later, when we consider
its role in the LET theory.

Leith's optimistic view not only inspired successful practical
applications of the FDR to study climate sensitivity \cite{Bell80} and
viscosity renormalization \cite{Carnevale83}, but was also seen as
seminal in stimulating an important series of papers which examined the
applicability of the FDR from the point of view of dynamical systems
theory \cite{Falcioni90} - \nocite{Carnevale91}
\cite{Boffetta03}. The overall conclusion of these papers can be stated
as follows:
\begin{enumerate}
\item A general relationship exists for the response of a chaotic system
in terms of its stationary probability distribution provided that the
system is dynamically mixing.
\item If the stationary probability is Gaussian in form, then the
relationship reduces to the FDR as given by equation (\ref{fdr}).
\end{enumerate}
Of course in real fluid turbulence the probability distribution is not
Gaussian, nor is it known exactly. However, as we have shown in
\cite{Kiyani04}, the FDR can be derived for turbulence to second-order
in renormalized perturbation theory and hence, if used appropriately, is
consistent with a closure approximation of that order. We shall return
to this point later.

\subsection{The time-ordered FDR}

In \cite{Kiyani04} we postulated that in the context of the
Kraichnan-Wyld perturbation theory we may re-write the existing relationship
between the zero-order covariance and zero-order response in a \emph{renormalized} form
as
\beq
C_{\alpha\sigma}\left(\mathbf{k};t,t'\right)=\theta\left(t-s\right)R_{\alpha\epsilon}\left(\mathbf{k};t,s\right)
C_{\epsilon\sigma}\left(\mathbf{k};s,t'\right),
\eeq
or in its isotropic version as
\beq
C\left(k;t,t'\right)=\theta\left(t-s\right)R\left(k;t,s\right)C\left(k;s,t'\right),
\label{eq:PreLETAxiom}
\eeq
where the Heaviside unit-step function $\theta(t-s)$ explicitly states
the causality condition. 
As yet we have taken no decision about the ordering of the two times
$t$ and $t'$, and thus the symmetry under interchange of $t$ and
$t'$ is untested in (\ref{eq:PreLETAxiom}).

If we explicitly state the time ordering as $t>t'$ say, then this
is equivalent to applying $\theta(t-t')$
to both sides of (\ref{eq:PreLETAxiom}):
\beq
\theta\left(t-t'\right)C\left(k;t,t'\right)=\theta\left(t-t'\right)\theta\left(t-s\right)R\left(k;t,s\right)C\left(k;s,t'\right),
\label{eq:NewLETAxiom}
\eeq
and this is the beginning of the LET theory. In it, we have postulated
the existence of a renormalized propagator and have made 
use of the Heaviside unit-step function to make the time-ordering
manifest.

The generalized fluctuation dissipation relationship
is obtained by setting $s=t'$ in (\ref{eq:NewLETAxiom}) to
get 
\beq
\theta\left(t-t'\right)C\left(k;t,t'\right)=\theta\left(t-t'\right)
R\left(k;t,t'\right)C\left(k;t',t'\right),
\label{eq:FDT}
\eeq
where the time-ordering is set by the requirement $s=t'$. 

In \cite{Kiyani04} we introduced a representation of the covariance which
preserves its symmetry under
interchange of time arguments, thus:
\bea
C\left(k;t,t'\right) & = & \theta\left(t-t'\right)C\left(k;t,t'\right)+\theta\left(t'-t\right)
C\left(k;t,t'\right) \nonumber \\
& & -\delta_{t,t'}C\left(k;t,t'\right).
\label{eq:symCor0}
\eea
We can easily show that this representation does what it is supposed to do
by looking in turn at the separate cases: $t<t';\quad t>t'$ and $t=t'$, 
and this is left for the reader.

Now, using (\ref{eq:NewLETAxiom}) to expand the right hand side of (\ref{eq:symCor0})
we obtain
\begin{eqnarray}
C\left(k;t,t'\right) & = & \theta\left(t-t'\right)\theta\left(t-s\right)R\left(k;t,s\right)C\left(k;s,t'\right)\nonumber \\
 &  & +\theta\left(t'-t\right)\theta\left(t'-p\right)R\left(k;t',p\right)C\left(k;p,t\right)\nonumber \\
 &  & -\delta_{t,t'}C\left(k;t,t'\right).
 \label{eq:symCorrel}
 \end{eqnarray}
Or, this result may be written more like the FDR by instead using
(\ref{eq:FDT}) to construct it 
\begin{eqnarray}
C\left(k;t,t'\right) & = & \theta\left(t-t'\right)R\left(k;t,t'\right)C\left(k;t',t'\right)\nonumber \\
 &  & +\theta\left(t'-t\right)R\left(k;t',t\right)C\left(k;t,t\right)\nonumber \\
 &  & -\delta_{t,t'}C\left(k;t,t'\right).
 \label{eq:symFDT}
 \end{eqnarray}
The symmetry of both these covariances, (\ref{eq:symCorrel}) and
(\ref{eq:symFDT}), can be broken by applying a unit-step function
to both sides. This will yield either (\ref{eq:NewLETAxiom})
or (\ref{eq:FDT}), depending on which time-ordering we choose.
\section{The properties of the mean-field propagator}

In this section we begin by reviewing the introduction of a velocity
propagator, as in the original formulation of the LET theory
\cite{McComb78} and note that in this context the propagator introduced
in \cite{Kiyani04} is a mean-field propagator.

\subsection{The velocity field propagator}

From the exact solution of the solenoidal NSE (see \cite{McComb90a}), we have:
\begin{eqnarray}
u_{\alpha}\left(\mathbf{k},t\right) & = & \widehat{R}_{\alpha\sigma}^{(0)}
\left(\mathbf{k};t,s\right)
u_{\sigma}\left(\mathbf{k},s\right)+\nonumber \\
 &  & + \left[ \lambda \int_{s}^{t}dt''\widehat{R}_{\alpha\sigma}^{(0)}
 \left(\mathbf{k};t,t''\right) \right. \nonumber \\
& & \times \left. \int d^{3}jM_{\sigma\beta\gamma}(\mathbf{k})u_{\beta}\left(\mathbf{j},t''\right)
 u_{\gamma}\left(\mathbf{k-j},t''\right) \right] ,\nonumber \\
\label{eq:A2}
\end{eqnarray}
where $\widehat{R}_{\alpha\sigma}^{(0)}$ is the `viscous' or zero-order response tensor and the hat is used to emphasize that this is the `response' associated with the instantaneous velocity field.

Expanding $u_{\alpha}\left(\mathbf{k},t\right)$ in a perturbation
series around a gaussian solution and equating zero-order terms we
can say that the equality
\beq
u_{\alpha}^{(0)}\left(\mathbf{k},t\right)=\widehat{R}_{\alpha\sigma}^{(0)}
\left(\mathbf{k};t,s\right)u_{\alpha}^{(0)}\left(\mathbf{k},s\right),
\label{eq:A1}
\eeq
illustrates the propagator-like nature of $\widehat{R}_{\alpha\sigma}^{(0)}
\left(\mathbf{k};t,s\right)$.
Then from looking at the form of (\ref{eq:A2}), we can \emph{postulate}
the existence of a renormalized propagator such that we obtain
a renormalized version of (\ref{eq:A1})
\beq
u_{\alpha}\left(\mathbf{k},t\right)=\widehat{R}_{\alpha\sigma}
\left(\mathbf{k};t,s\right)u_{\alpha}\left(\mathbf{k},s\right).
\label{eq:postulate1}
\eeq
Multiply (\ref{eq:postulate1}) by $u_{\beta}\left(\mathbf{-k},t'\right)$
\beq
u_{\alpha}\left(\mathbf{k},t\right)u_{\beta}\left(\mathbf{-k},t'\right)=\widehat{R}_{\alpha\sigma}\left(\mathbf{k};t,s\right)u_{\sigma}
\left(\mathbf{k},s\right)u_{\beta}\left(\mathbf{-k},t'\right),
\eeq
and average this equation to obtain
\beq
C_{\alpha\beta}(\mathbf{k};t,t')=R_{\alpha\sigma}\left(\mathbf{k};t,s\right)
C_{\sigma\beta}\left(\mathbf{k};s,t'\right),
\label{eq:postulate2}
\eeq
where the propagator is statistically independent of the velocity
field and we have used the mean-field approximation
\beq
\left\langle \widehat{R}_{\alpha\sigma}\left(\mathbf{k};t,s\right)\right\rangle =R_{\alpha\sigma}\left(\mathbf{k};t,s\right).
\label{eq:MeanField}
\eeq
As usual, equation (\ref{eq:postulate2}) can be turned into a simpler scalar form by
using the properties of isotropic tensors
\beq
C\left(k;t,t'\right)=\theta\left(t-s\right)R\left(k;t,s\right)C\left(k;s,t'\right).
\label{eq:govneqn}
\eeq

The transitivity of $\hat{R}_{\alpha r}(\mathbf{k};t,s)$ with respect to
intermediate time can be proved by applying equation (\ref{eq:postulate1}) to the
right-hand side of itself
\beq
u_{\alpha}\left(\mathbf{k},t\right)=\widehat{R}_{\alpha\sigma}
\left(\mathbf{k};t,s\right)\widehat{R}_{\sigma\rho}\left(\mathbf{k};s,t'\right)
u_{\rho}\left(\mathbf{k},t'\right),
\eeq
and realising that we could also have written this as
\beq
u_{\alpha}\left(\mathbf{k},t\right)=\widehat{R}_{\alpha\rho}\left(\mathbf{k};t,t'\right)
u_{\rho}\left(\mathbf{k},t'\right),
\eeq
implying the result:
\beq
\widehat{R}_{\alpha\rho}\left(\mathbf{k};t,t'\right)=\widehat{R}_
{\alpha\sigma}\left(\mathbf{k};t,s\right)\widehat{R}_{\sigma\rho}
\left(\mathbf{k};s,t'\right) ,
\eeq
and $t>s>t'$.

\subsection{The mean-field propagator}

The simple property of the propagator
\beq
R(k;t,t)=1,
\eeq
 can be easily shown to be necessary by setting $s=t$ in 
 (\ref{eq:NewLETAxiom}).
Other properties can be obtained by equating the right hand side 
of (\ref{eq:NewLETAxiom})
with the right hand side of (\ref{eq:FDT}):
\begin{eqnarray}
& & \theta\left(t-t'\right)R\left(k;t,t'\right)C\left(k;t',t'\right) \nonumber \\
& & = \theta\left(t-t'\right)\theta\left(t-s\right)R\left(k;t,s\right)C\left(k;s,t'\right).
\label{eq:fdt2}
\end{eqnarray}
Expanding the right hand side of (\ref{eq:fdt2}) using (\ref{eq:symFDT}) we
obtain
\begin{eqnarray}
 &  & \theta\left(t-t'\right)R\left(k;t,t'\right)C\left(k;t',t'\right)\nonumber \\
 &  & \left.\begin{array}{c}
=\left[\theta\left(t-t'\right)\theta\left(t-s\right)R\left(k;t,s\right)\times\right.\\
\left.\times\theta\left(s-t'\right)R\left(k;s,t'\right)C\left(k;t',t'\right)\right]
\end{array}\right\} a\nonumber \\
 &  & \left.\begin{array}{c}
+\left[\theta\left(t-t'\right)\theta\left(t-s\right)R\left(k;t,s\right)\right.\times\\
\times\left.\theta\left(t'-s\right)R\left(k;t',s\right)C\left(k;s,s\right)\right]
\end{array}\right\} b\nonumber \\
 &  & \left.\begin{array}{c}
-\left[\theta\left(t-t'\right)\theta\left(t-s\right)R\left(k;t,s\right)\right.\\
\times\left.\delta_{t',s}C\left(k;s,t'\right)\right].\end{array}\right\} c
\label{eq:transprf1}
\end{eqnarray}
 Dividing the right hand side into three groups of terms labelled respectively
$a$,$b$ and $c$, we will now look at (\ref{eq:transprf1}) for
the two separate cases:- \textbf{Case} 1, $t>s>t'$; and \textbf{Case} 2 $t\geq t'>s$.
\\
\subsubsection{Transitivity with respect to intermediate times}

Here we consider case 1 corresponding to $t>s>t'$ and
implying that $b=0$ and $c=0$ in equation (\ref{eq:transprf1}). This will leave
\begin{eqnarray}
 &  & \theta\left(t-t'\right)R\left(k;t,t'\right)C\left(k;t',t'\right)\nonumber \\
 &  & =\left[\theta\left(t-t'\right)\theta\left(t-s\right)R\left(k;t,s\right)
 \times\right.\nonumber \\
 &  & \left.\times\theta\left(s-t'\right)R\left(k;s,t'\right)C\left(k;t',t'\right)\right].
 \label{eq:transprf2}
 \end{eqnarray}
We now use the contraction property of the Heaviside function 
\beq
\theta\left(t-s\right)\theta\left(s-t'\right)=\theta\left(t-t'\right),
\label{eq:thetacontr34}
\eeq
to write (\ref{eq:transprf2}) as
\begin{eqnarray}
 &  & \theta\left(t-t'\right)\underline{R\left(k;t,t'\right)}C\left(k;t',t'\right)
 \nonumber \\
 &  & =\theta\left(t-t'\right)\underline{R\left(k;t,s\right)R\left(k;s,t'\right)}
 C\left(k;t',t'\right).
 \end{eqnarray}
From this above result, we can deduce the \emph{transitive} property
of the propagator
\beq
R\left(k;t,t'\right)=R\left(k;t,s\right)R\left(k;s,t'\right).
\label{eq:transitive}
\eeq 
This result also tells us that the transitivity of the propagator
holds only for times $s$ which are \emph{intermediate} between the
two times $t$ and $t'$. This makes sense because otherwise, if $s$
was outside the range between $t$ and $t'$, we would have propagation
backwards in time which violates causality.
This is a result which was previously only assumed \cite{McComb78},
\cite{Oberlack01} on the basis that it
could be expected to follow from the corresponding relationship for the
velocity-field propagator, and is now proved.

\subsubsection{Linked single-time covariances}

Next we consider
\textbf{\emph{Case 2}} $t\geq t'>s$, which
corresponds to $a=0$ and $c=0$, leaving
\begin{eqnarray}
 &  & \theta\left(t-t'\right)R\left(k;t,t'\right)\underline{C\left(k;t',t'\right)}\nonumber \\
 &  & =\left[\theta\left(t-t'\right)\theta\left(t-s\right)R\left(k;t,s\right)\right.\nonumber \\
 &  & \times\left.\theta\left(t'-s\right)R\left(k;t',s\right)\underline{C\left(k;s,s\right)}\right]
 \end{eqnarray}
This result is important because it links two single-time covariances.
This fact becomes clearer if we take the special case of $t=t'$.
This gives
\beq
C\left(k;t,t\right)=\theta\left(t-s\right)R\left(k;t,s\right)R\left(k;t,s\right)C\left(k;s,s\right),
\label{eq:1tLink}
\eeq
implying that we need two propagators to link single-time covariances.
Defining 
\beq
\widetilde{R}\left(k;t,s\right):=R\left(k;t,s\right)R\left(k;t,s\right),
\label{eq:LinkProp}
\eeq
Equation (\ref{eq:1tLink}) can be modified to make it look like equation (\ref{eq:PreLETAxiom})
 \beq
C\left(k,t\right)=\theta\left(t-s\right)\widetilde{R}\left(k;t,s\right)C\left(k,s\right).
\label{eq:LinkEqn}
\eeq
Again, the presence of the unit-step function ensures that the covariance
can only propagate forwards in time. 

\section{Derivation of the local energy transfer (LET) response equation}

The starting point for the LET theory is the second-order
renormalized covariance equation as given by (\ref{dia}).
We can now proceed in two ways. 
\begin{enumerate}
\item The first is to substitute (\ref{eq:symFDT}) in (\ref{dia})
and then choose $t>t'$.
\item The second is to choose $t>t'$ and multiply both sides of 
(\ref{dia})
by $\theta\left(t-t'\right)$ to show the range over which the equation
will be valid. Then follow this by using the FDR, in the form of equation
(\ref{eq:FDT}), throughout. 
\end{enumerate}
\begin{quote}
\emph{Note} it is important that we do not set $t=t'$
in the covariance equation (\ref{dia}) as we can only do this after evaluating
the derivative of the 2-time covariance. 
\end{quote}
Both methods are equivalent but the second is the easier to use in
practice. Thus we begin by choosing the time-ordering to be $t\geq t'$
and multiplying (\ref{dia}) by $\theta(t-t')$ 
\begin{eqnarray}
 &  & \theta(t-t')\frac{\partial}{\partial t}C\left(k;t,t'\right)+\theta(t-t')\nu k^{2}C
 \left(k;t,t'\right)\nonumber \\
 &  & =\theta(t-t')\int d^{3}jL\left(\mathbf{k},\mathbf{j}\right) \nonumber \\ 
 & & \times \left\{ \int_{0}^{t'}dsR\left(k;t',s\right)C\left(j;t,s\right)C\left(\left|\mathbf{k}-\mathbf{j}\right|;t,s\right)\right.\nonumber \\
 &  & \left.-\int_{0}^{t}dsR\left(j;t,s\right)C\left(k;s,t'\right)C\left(\left|\mathbf{k}-\mathbf{j}\right|;t,s\right)\right\} .
 \label{eq:BrokenCov}
 \end{eqnarray}

Let us look at the first term of the left hand side of (\ref{eq:BrokenCov}):
\begin{eqnarray}
 &  & \theta\left(t-t'\right)\frac{\partial}{\partial t}
 C\left(k;t,t'\right)\nonumber \\
 &  & =\frac{\partial}{\partial t}\theta\left(t-t'\right)C\left(k;t,t'\right)-
 C\left(k;t,t'\right)\frac{\partial}{\partial t}\theta\left(t-t'\right)\nonumber \\
 &  & =\frac{\partial}{\partial t}\theta\left(t-t'\right)R\left(k;t,t'\right)
 C\left(k;t',t'\right)\nonumber \\
 &  & \,\,\,-C\left(k;t,t'\right)\frac{\partial}{\partial t}\theta\left(t-t'\right),
 \end{eqnarray}
where we have applied the product rule in the second line, and
the FDT, (\ref{eq:FDT}) in the third line. After substituting
the differential of the Heaviside unit-step function
\beq
\frac{\partial}{\partial t}\theta\left(t-t'\right)=\delta(t-t'),
\eeq
 we reach our final form for this part of the response equation. Thus:
 \begin{eqnarray}
 &  & \!\!\!\!\!
 \mbox{left hand side}\,\,\mbox{of}\,\,(\ref{eq:BrokenCov})\nonumber \\
 &  & \!\!\!\!\!
 =\frac{\partial}{\partial t}\theta\left(t-t'\right)R\left(k;t,t'\right)C\left(k;t',t'\right)-C\left(k;t,t'\right)\delta(t-t')\nonumber \\
 &  & \!\!\!\!\!
 +\nu k^{2}\theta\left(t-t'\right)R\left(k;t,t'\right)C\left(k;t',t'\right),
 \end{eqnarray}
where the FDR (\ref{eq:FDT}) was used on the second term of the left
hand side
of (\ref{eq:BrokenCov}) also.

Now we evaluate the second time integral on the right hand side of 
(\ref{eq:BrokenCov}), which we label as $TI_{2}$:
\beq
TI_{2}=\theta\left(t-t'\right)\int_{0}^{t}dsR\left(j;t,s\right)C\left(k;s,t'\right)
C\left(\left|\mathbf{k}-\mathbf{j}\right|;t,s\right).
\label{eq:expaintegral}
\eeq

We need to have the appropriate $\theta$ functions in front of the
covariance so that the broken time-reversal symmetry becomes
manifest. This information is present in the arguments of the propagator
and in $\theta\left(t-t'\right)$. 
So for $C\left(\left|\mathbf{k}-\mathbf{j}\right|;t,s\right)$
\begin{eqnarray}
& & \theta\left(t-t'\right)\int_{0}^{t}dsC\left(\left|\mathbf{k}-\mathbf{j}\right|;t,s\right) \nonumber \\
& & = \theta\left(t-t'\right)\int_{0}^{t}ds\theta\left(t-s\right)
C\left(\left|\mathbf{k}-\mathbf{j}\right|;t,s\right),
\end{eqnarray}
 and for $C\left(k;s,t'\right)$
 \begin{eqnarray}
& & \theta\left(t-t'\right)\int_{0}^{t}ds C\left(k;s,t'\right) \nonumber \\
& & = \theta\left(t-t'\right)\int_{0}^{t'}ds
C\left(k;s,t'\right)\nonumber \\
 & & + \theta\left(t-t'\right)\int_{t'}^{t}dsC\left(k;s,t'\right)\nonumber \\
 & & = \theta\left(t-t'\right)\int_{0}^{t'}ds\theta\left(t'-s\right)
 C\left(k;t',s\right)\nonumber \\
 & & + \theta\left(t-t'\right)\int_{t'}^{t}ds\theta\left(s-t'\right)
 C\left(k;s,t'\right),
 \end{eqnarray}
where we have used the property $C(k;t,t') = C(k;t',t)$ in the fourth line. With these
results we can now write (\ref{eq:expaintegral}) as 
\begin{eqnarray}
 &  & TI_{2}\nonumber \\
 &  & =\left[\theta\left(t-t'\right)\int_{t'}^{t}dsR\left(j;t,s\right)
 \theta\left(s-t'\right)\times\right.\nonumber \\
 &  & \left.\frac{}{}\times C\left(k;s,t'\right)\theta\left(t-s\right)
 C\left(\left|\mathbf{k}-\mathbf{j}\right|;t,s\right)\right]\nonumber \\
 &  & +\left[\theta\left(t-t'\right)\int_{0}^{t'}dsR\left(j;t,s\right)\theta
 \left(t'-s\right)\times\right.\nonumber \\
 &  & \left.\frac{}{}\times C\left(k;t',s\right)\theta\left(t-s\right)
 C\left(\left|\mathbf{k}-\mathbf{j}\right|;t,s\right)\right].
 \end{eqnarray}

 The evaluation of the first integral on the right hand side of (\ref{eq:BrokenCov})
follows similarly so that the final LET response equation is given
by
\begin{eqnarray}
 &  & \frac{\partial}{\partial t}\theta\left(t-t'\right)R\left(k;t,t'\right)C\left(k;t',t'\right)-C\left(k;t,t'\right)\delta\left(t-t'\right)\nonumber \\
 &  & +\nu k^{2}\theta\left(t-t'\right)R\left(k;t,t'\right)C\left(k;t',t'\right)\nonumber \\
 & & = \int d^{3}jL\left(\mathbf{k},\mathbf{j}\right)\theta\left(t-t'\right)\times\nonumber \\
 &  & \times\left\{ \left[ \int_{0}^{t'}dsR\left(k;t',s\right)\theta\left(t-s\right)\times \right. \right. \nonumber \\ & &  \left. \frac{}{} \times C\left(j;t,s\right)\theta\left(t-s\right)C\left(\left|\mathbf{k}-\mathbf{j}\right|;t,s\right)\right] \nonumber \\
 &  & - \left[ \int_{0}^{t'}dsR\left(j;t,s\right)\theta\left(t'-s\right) \times \right. \nonumber \\
 & & \left. \frac{}{} \times C\left(k;t',s\right)\theta\left(t-s\right)C\left(\left|\mathbf{k}-\mathbf{j}\right|;t,s\right ) \right] \nonumber \\
 &  & -\left[\int_{t'}^{t}dsR\left(j;t,s\right)\theta\left(s-t'\right)\times\right.\nonumber \\
 &  & \left.\left.\frac{}{}\times C\left(k;s,t'\right)\theta\left(t-s\right)C\left(\left|\mathbf{k}-\mathbf{j}\right|;t,s\right)\right]\right\} .
 \end{eqnarray}
Multiplying both sides by $\theta(t-t')$, dividing by $C\left(k;t',t'\right)$
and noting that 
\beq
\frac{\theta(t-t')C\left(k;t,t'\right)}{C\left(k;t',t'\right)}=\theta(t-t')R(k;t,t'),
\eeq
from the FDR, in the form of equation (\ref{eq:FDT}), we reach the simplified form with the
broken time-reversal symmetry manifest
\begin{eqnarray}
 &  & \theta\left(t-t'\right)\left(\frac{\partial}{\partial t}+\nu k^{2}\right)
 \theta\left(t-t'\right)R\left(k;t,t'\right) \nonumber \\
 & & -\theta(t-t')R(k;t,t')\delta\left(t-t'\right)
 +\left[ \int d^{3}jL\left(\mathbf{k},\mathbf{j}\right)\theta\left(t-t'\right) \right. \nonumber \\
 &  & \times \left. \int_{t'}^{t}dsR\left(j;t,s\right)R\left(k;s,t'\right)\theta\left(t-s\right)
 C\left(\left|\mathbf{k}-\mathbf{j}\right|;t,s\right) \right] \nonumber \\
 &  & =\int d^{3}jL\left(\mathbf{k},\mathbf{j}\right)\theta\left(t-t'\right)
 \int_{0}^{t'}ds\frac{\theta\left(t-s\right)C\left(\left|\mathbf{k}-\mathbf{j}\right|;t,s\right)}{C\left(k;t',t'\right)} \nonumber \\
 &  & \times\left\{ R\left(k;t',s\right)\theta\left(t-s\right)C\left(j;t,s\right) \right. \nonumber \\
 & & - \left. R\left(j;t,s\right)\theta\left(t'-s\right)C\left(k;t',s\right)\right\} .
 \label{eq:LETresponse}
 \end{eqnarray}

\subsection{Comparison with previous forms}

Apart from the addition of the second term on the left hand side
\beq
-\theta(t-t')R(k;t,t')\delta\left(t-t'\right),
\eeq
equation (\ref{eq:LETresponse}) is the same as the LET response equation
which appears as equation (3.19) in \cite{McComb92},
equation (20) in \cite{McComb03} and equation (7.146) in \cite{McComb90a}. The natural
addition of this extra term as a consequence of time-ordering, fixes
the problem of the singularity in the time-derivative of the response
equation (\ref{eq:LETresponse}) which occurs when one takes $t=t'$.
More important, if we compare (64) with the DIA response equation (20),
the additional terms on the right hand side of (64) cancel the infra-red
divergence and ensure compatibility with the Kolmogorov K41 spectrum.

\section{The two-time LET theory}

\subsection{Partial-propagator representation}

We may write the propagator in a representation which separates the
discontinuous part as a Heaviside unit step function, thus: 
\beq
R(k;t,t')=\theta(t-t')\mathcal{R}(k;t,t'),
\label{eq:responseRep}
\eeq
where $\mathcal{R}(k;t,t')$ is a representation of the propagator
but without the discontinuity at $t=t'$. So using (\ref{eq:responseRep})
and the FDR (\ref{fdr}) to turn two-time covariances into single-time
form, (64) for the response function becomes:
\begin{eqnarray}
 &  & \!\!\!\!\!
 \theta\left(t-t'\right)\left(\frac{\partial}{\partial t}+\nu k^{2}\right)
 \mathcal{R}\left(k;t,t'\right)\nonumber \\
 &  & \!\!\!\!\!
 =-\theta\left(t-t'\right)\int d^{3}jL\left(\mathbf{k},\mathbf{j}\right)
\left[  \int_{t'}^{t}ds\mathcal{R}\left(k;s,t'\right) \times \right. \nonumber \\
 & & \left. \frac{}{} \times \mathcal{R}\left(j;t,s\right)
 \mathcal{R}\left(\left|\mathbf{k}-\mathbf{j}\right|;t,s\right)
 C\left(\left|\mathbf{k}-\mathbf{j} \right|,s\right) \right] \nonumber \\
 &  & \!\!\!\!\!
 +\theta\left(t-t'\right)\int d^{3}jL\left(\mathbf{k},\mathbf{j}
 \right)\int_{0}^{t'}ds\left\{ \frac{}{}\mathcal{R}\left(k;t',s\right)
 \mathcal{R}\left(j;t,s\right)
\times \right.\nonumber \\
 &  & \!\!\!\!\!\!\!\!\!\!
 \left.\times \mathcal{R}\left(\left|\mathbf{k}-\mathbf{j}\right|;t,s\right)
 \frac{C\left(\left|\mathbf{k}-\mathbf{j}\right|,s\right)}
 {C\left(k,t'\right)}\left[C\left(j,s\right)-C\left(k,s\right)\right]\right\} ,
 \label{eq:response03}
 \end{eqnarray}
 for $t\geq t'$. The counter-term has been
cancelled by use of the product rule in the time-derivative.

\subsection{The LET closure equations} 
 
The LET Equations may now be summarised as follows. For the two-time covariance, we have equation (\ref{dia}), 
\begin{eqnarray}
 &  & \!\!\!\!\!\!\!\!\!\!\!\!\!\!\!
 \left(\frac{\partial}{\partial t}+\nu k^{2}\right)C\left(k;t,t'\right)=\int d^{3}jL\left(\mathbf{k},\mathbf{j}\right)\times\nonumber \\
 &  & \!\!\!\!\!
 \times\left\{ \int_{0}^{t'}dsR\left(k;t',s\right)C\left(j;t,s\right)C\left(\left|\mathbf{k}-\mathbf{j}\right|;t,s\right)\right.\nonumber \\
 &  & \!\!\!\!\!
 \left.-\int_{0}^{t}dsR\left(j;t,s\right)C\left(k;s,t'\right)C\left(\left|\mathbf{k}-\mathbf{j}\right|;t,s\right)\right\} ,
 \label{eq:2tcor}
 \end{eqnarray}
and likewise equation (\ref{diag}) for the single-time covariance:
 \begin{eqnarray}
 &  & \!\!\!\!\!\!\!\!\!\!\!\!\!\!\!
 \left(\frac{\partial}{\partial t}+2\nu k^{2}\right)C\left(k,t\right)=2\int d^{3}jL\left(\mathbf{k},\mathbf{j}\right)\times \nonumber \\
&  & \!\!\!\!\!
\times\int_{0}^{t}dsR\left(k;t,s\right)
R\left(j;t,s\right)R\left(\left|\mathbf{k}-\mathbf{j}\right|;t,s\right)\times \nonumber \\
&  & \!\!\!\!\!
 \left[C\left(j,s\right)C\left(\left|\mathbf{k}-\mathbf{j}\right|,s\right)-C\left(k,s\right)C\left(\left|\mathbf{k}-\mathbf{j}\right|,s\right)\right].
 \label{eq:1-tcor}  
 \end{eqnarray}
where we have invoked the FDR so that all two-time covariances
are turned into one-time covariances.

For the response function we can use either (64) or (67). The above
equations along with the generalised
fluctuation-dissipation relation (FDR),
\beq
\theta(t-t')C(k;t,t')=\theta(t-t')R(k;t,t')C(k,t'),
\label{eq:FDR}
\eeq
from which the LET is derived, and the single-time covariance link equation
\beq
C(k,t)=\theta(t-s)R(k;t,s)R(k;t,s)C(k,s),
\label{eq:linkEqn}
\eeq
complete the set of LET equations.

The LET equations have been applied, along with those of the DIA, to the
problem of free decay of isotropic turbulence from arbitrary initial
conditions, over a wide range of Taylor-Reynolds numbers
\cite{McComb84}, \cite{McComb89}. In these investigations, the
covariance equations on and off the time diagonal were solved
simultaneously with the relevant response equation. It was later
realized that for the LET theory, the response equation could be
replaced by the FDR, as given by (70), and this reduced the
computational effort well below that of DIA: see \cite{McComb92},
\cite{McComb03}. This work was for three-dimensional turbulence, while
an extensive investigation of the two-dimensional case has also been
carried out for DIA, SCF and LET theories \cite{Frederiksen94},
\cite{Frederiksen97a}, \cite{Frederiksen00}.

\subsection{Behaviour in the limit of infinite Reynolds numbers}

The later two-time versions of the LET theory claimed that their solutions
were compatible with K41. However, this was never shown explicitly.
Compatibility with K41 is now demonstrated for the LET response/propagator
equation as given by (\ref{eq:response03}). 
We begin by writing (\ref{eq:response03}) in stationary form. This
means that all (single-time) covariances become time independent:
\beq
C(k,t)\rightarrow C(k),
\eeq
and we write the propagator in relative time coordinates
\beq
\mathcal{R}\left(k;t,t'\right)=\mathcal{R}\left(k;t-t'\right).
\eeq
Next we assume the exponential form for the propagator
\beq
\mathcal{R}\left(k;t-t'\right)=\exp[-\omega(k)(t-t')],
\eeq
where, as before, $\omega(k)$ is the total eddy-decay rate. These
changes result in the response equation becoming:
\begin{eqnarray}
 &  & \!\!\!\!\!\!\!\!\!\!
 \theta\left(t-t'\right)\left(\frac{\partial}{\partial t}+\nu k^{2}\right)
 \exp[-\omega(k)(t-t')]\nonumber \\
 &  & \!\!\!\!\!\!\!\!\!\!
 =\left[-\theta\left(t-t'\right)\int d^{3}jL\left(\mathbf{k},\mathbf{j}\right)
 C\left(\left|\mathbf{k}-\mathbf{j}\right|\right)\int_{t'}^{t}ds\times\right.\nonumber \\
 &  & \!\!\!\!\!\!\!\!\!\!
 \left.\frac{}{}\times\exp[-\omega(k)(s-t')-\omega(j)(t-s)-
 \omega(\left|\mathbf{k}-\mathbf{j}\right|)(t-s)]\right]\nonumber \\
 &  & \!\!\!\!\!\!\!\!\!\!
 +\left[\theta\left(t-t'\right)\int d^{3}jL\left(\mathbf{k},
 \mathbf{j}\right)\int_{0}^{t'}ds\times\right.\nonumber \\
 &  & \!\!\!\!\!\!\!\!\!\!
 \times\left\{ \frac{}{}\exp[-\omega(k)(t'-s)-
 \omega(j)(t-s)-\omega(\left|\mathbf{k}-\mathbf{j}\right|)(t-s)]
\right.\nonumber \\
 &  & \!\!\!\!\!\!\!\!\!\!
 \left.\left.\times\frac{C\left(\left|\mathbf{k}-\mathbf{j}\right|\right)}
 {C\left(k\right)}\left[C\left(j\right)-C\left(k\right)\right]\right\} \right].
 \end{eqnarray}
Doing the differentiation, setting $t=t'$ and carrying out the time integration
results in an equation for $\omega(k)$:
\begin{eqnarray}
& & \!\!\!\!\!\!\!\!\!\!
\omega(k)  =  \nu k^{2}\nonumber \\
& &+ \left\{ \int d^{3}jL\left(\mathbf{k},
\mathbf{j}\right)\frac{C\left(\left|\mathbf{k}-\mathbf{j}\right|\right)\left[C\left(j\right)-C\left(k\right)\right]}{C\left(k\right)\left[\omega(k)+\omega(j)+\omega(\left|\mathbf{k}-\mathbf{j}\right|)\right]}\times\right.\nonumber \\
 &  & \left.\times\frac{}{}(1-\exp[-(\omega(k)+\omega(j)+\omega(\left|\mathbf{k}-\mathbf{j}\right|))t])\right\} ,
 \label{eq:StatResponse}
 \end{eqnarray}
where one can ignore the last term involving the exponential as we
are considering stationary systems which are time independent. Another
way to justify the neglect of this term is to realise that it
originates from the fact that we chose to have the initial conditions
at $t=0$ rather than the more usual $t=-\infty$.

To show that (\ref{eq:StatResponse}) is not divergent we complete
our analysis by substituting the inertial range/ infinite Reynolds
number forms for $C(k)$ and $\omega(k)$: 
\beq
C(k)=\frac{\alpha\varepsilon^{2/3}}{4\pi}k^{-11/3},
\eeq
\beq
\omega(k)=\beta\varepsilon^{1/3}k^{2/3},
\eeq
where $\alpha$ and $\beta$ are constants,
and by writing the integral in $k,j,\mu$ variables
\begin{eqnarray}
& &  \!\!\!\!\!
\omega(k) = \nu k^{2}\nonumber \\
& & +\left\{ \int dj\int d\mu\frac{kj^{3}(\mu^{2}-1)
[\mu(k^{2}+j^{2})-kj(1+2\mu^{2})]}{k^{2}+j^{2}-2kj\mu}
\times\right.\nonumber \\
 &  & \left.\times\frac{\alpha\beta^{-1}\varepsilon^{1/3}
 \left|\mathbf{k}-\mathbf{j}\right|^{-11/3}\left[j^{-11/3}-k^{-11/3}\right]}
 {k^{-11/3}\left[k^{2/3}+j^{2/3}+\left|\mathbf{k}-\mathbf{j}\right|^{2/3}\right]}\right\} 
 \label{eq:LimInfReyn}
 \end{eqnarray}
where $\mu$ is the cosine of the angle between the two vectors $\mathbf{k}$
and $\mathbf{j}$. 

There are three possible sources of divergence (of the infra-red type)
in this expression. However, from equation (\ref{eq:LimInfReyn}), it may be seen that the $k\rightarrow0$
and $j\rightarrow0$ limits do not pose a problem. The final possible
source of trouble $\left|\mathbf{k}-\mathbf{j}\right|\rightarrow0$ can be resolved
by realising that the term $\left[j^{-11/3}-k^{-11/3}\right]$cancels
the divergence caused by the $\left|\mathbf{k}-\mathbf{j}\right|^{-11/3}$
term. This is shown by expanding 
\beq
\left|\mathbf{k}-\mathbf{j}\right|^{-11/3}=\left(k^{2}+j^{2}-2kj\mu\right)^{-11/6},
\eeq
and substituting in equation (\ref{eq:LimInfReyn}). One then Taylor
expands $k$ around $j$ to leading order in $\epsilon=k-j$ in both
the numerator and denominator of the integrand in equation 
(\ref{eq:LimInfReyn}). This results in the integrand becoming
\bea
& & \!\!\!\!\!\!\!\!\!\!\!\!
\frac{(11/3)\alpha\beta^{-1}\varepsilon^{1/3}}{(2)^{17/6}} \times \nonumber \\
& & \times \frac{\left[2j^{2}\mu(1-\mu)-j^{2}\right](\mu^{2}-1)(1-\mu)^{-17/6}\epsilon}{j^{16/6}
\left[2j^{2/3}+(2j^{2})^{1/3}(1-\mu)^{1/3}\right]}\;,
\label{eq:LimInfReynB}
\eea
and focusing on the term $(\mu^{2}-1)(1-\mu)^{-17/6}\epsilon$ we
can see that as $\epsilon\rightarrow0$, the integrand goes to zero,
except at $\mu=1$ where the integrand is singular. This singularity
can be avoided if we write the limits of the $\mu$ integral as
\beq
\int_{-1}^{1}d\mu\rightarrow\int_{-1}^{\uparrow1}d\mu\;,
\eeq
where $\uparrow1$ implies in the limit approaching $1$ from below.

This completes the analysis in the limit of infinite Reynolds number.
Further information on the above technique can be found in \cite{McComb90a}.

\section{Single-time Markovianized LET theory}

The relevant single time LET equations are the single-time covariance
(using the partial propagator form):
\begin{eqnarray}
 &  & \!\!\!\!\!\!\!\!\!\!
  \left(\frac{\partial}{\partial t}+2\nu k^{2}\right)C\left(k,t\right)
 =2\int d^{3}jL\left(\mathbf{k},\mathbf{j}\right) \times \nonumber \\
 &  & \!\!\!\!\!\!\!\!\!\!
  \times \int_{0}^{t}ds\mathcal{R}
 \left(k;t,s\right)\mathcal{R}\left(j;t,s\right)\mathcal{R}\left(\left|\mathbf{k}-\mathbf{j}\right|;t,s\right)\times\nonumber \\
 &  & \!\!\!\!\!\!\!\!\!\!
  \left[C\left(j,s\right)C\left(\left|\mathbf{k}-\mathbf{j}\right|,s\right)-C\left(k,s\right)C\left(\left|\mathbf{k}-\mathbf{j}\right|,s\right)\right],
 \label{eq:1-tcor2}
 \end{eqnarray}
the response equation
\begin{eqnarray}
 &  &  \!\!\!\!\!
  \theta\left(t-t'\right)\left(\frac{\partial}{\partial t}+\nu k^{2}\right)\mathcal{R}\left(k;t,t'\right)\nonumber \\
 &  & \!\!\!\!\!
  =-\theta\left(t-t'\right)\int d^{3}jL\left(\mathbf{k},\mathbf{j}\right)
\left[  \int_{t'}^{t}ds \mathcal{R}\left(k;s,t'\right) \right. \times \nonumber \\
 & & \!\!\!\!\!
  \left. \times \frac{}{} \mathcal{R}\left(j;t,s\right)
 \mathcal{R}\left(\left|\mathbf{k}-\mathbf{j}\right|;t,s\right)
 C\left(\left|\mathbf{k}-\mathbf{j}\right|,s\right) \right] \nonumber \\
 &  & \!\!\!\!\!
  +\theta\left(t-t'\right)\int d^{3}jL\left(\mathbf{k},\mathbf{j}\right)
 \int_{0}^{t'}ds\left\{ \frac{}{}\mathcal{R}\left(k;t',s\right)\mathcal{R}
 \left(j;t,s\right)\times\right.\nonumber \\
 &  & \!\!\!\!\!\!\!\!\!\!
  \left.\times\mathcal{R}\left(\left|\mathbf{k}-\mathbf{j}\right|;t,s\right)
 \frac{C\left(\left|\mathbf{k}-\mathbf{j}\right|,s\right)}
 {C\left(k,t'\right)}\left[C\left(j,s\right)-C\left(k,s\right)\right]\right\} ,
 \label{eq:response04}
 \end{eqnarray}
and the single-time covariance link equation
\beq
C(k,t)=\theta(t-s)R(k;t,s)R(k;t,s)C(k,s),
\label{eq:Corlink}
\eeq

Making a Markovian approximation we can write (\ref{eq:1-tcor2})
as:
\bea
& &\!\!\!\!\!\!\!\!\!\!
 \left(\frac{\partial}{\partial t}+2\nu k^{2}\right)C\left(k,t\right)=2\int d^{3}jL\left(\mathbf{k},\mathbf{j}\right) D(k,j,\left|\mathbf{k}-
\mathbf{j}\right|;t) \nonumber \\
& & \quad \quad
 \times C\left(\left|\mathbf{k}-\mathbf{j}\right|,t\right)
\left[C\left(j,t\right)-C\left(k,t\right)\right],
\label{eq:Mark1tcor}
\eea
where the Markovian approximation amounts to updating each $C(s)$
to $C(t)$, and where
\bea
& & \!\!\!\!\!\!\!\!\!\!
D(k,j,\left|\mathbf{k}-\mathbf{j}\right|;t) \nonumber \\
& & =\int_{0}^{t}ds\mathcal{R}\left(k;t,s\right)
\mathcal{R}\left(j;t,s\right)\mathcal{R}\left(\left|\mathbf{k}-
\mathbf{j}\right|;t,s\right),
\label{eq:MemTime01}
\eea
is the \emph{memory time}. 

We now need some way of computing
$D(k,j,\left|\mathbf{k}-\mathbf{j}\right|;t)$; that is,
of updating it. We do this by differentiating (\ref{eq:MemTime01})
with respect to $t$
\begin{eqnarray}
& & \!\!\!\!\!\!
\frac{\partial}{\partial t}D(k,j,\left|\mathbf{k}-\mathbf{j}\right|;t) \nonumber \\
& & \!\!\!\!\!\!
=1 +\int_{0}^{t}ds\left[\left(\frac{\partial}{\partial t}\mathcal{R}\left(k;t,s\right)\right)\mathcal{R}\left(j;t,s\right)\mathcal{R}
\left(\left|\mathbf{k}-\mathbf{j}\right|;t,s\right)\right.\nonumber \\
 &  & \!\!\!\!\!\!
  +\mathcal{R}\left(k;t,s\right)\left(\frac{\partial}{\partial t}\mathcal{R}\left(j;t,s\right)\right)\mathcal{R}\left(\left|\mathbf{k}-\mathbf{j}
 \right|;t,s\right)\nonumber \\
 &  & \!\!\!\!\!\!
  \left.+\mathcal{R}\left(k;t,s\right)\mathcal{R}\left(j;t,s\right)
 \left(\frac{\partial}{\partial t}\mathcal{R}\left(\left|\mathbf{k}-\mathbf{j}
 \right|;t,s\right)\right)\right].
 \label{eq:memtdyneqn01}
 \end{eqnarray}

To evaluate (\ref{eq:memtdyneqn01}) we need to know the dynamical
behaviour of $\mathcal{R}\left(k;t,s\right)$. We obtain this from equation
(\ref{eq:response04}). Proceed by writing (\ref{eq:response04})
in Langevin form
\beq
\theta(t-t')\left[\frac{\partial}{\partial t}+\nu k^{2}+\eta(k;t,t')\right]
\mathcal{R}\left(k;t,t'\right)=0,
\label{eq:PropLang}
\eeq
where 
\begin{eqnarray}
 &  & \eta(k;t,t')\nonumber \\
 &  & =\theta\left(t-t'\right)\int d^{3}jL\left(\mathbf{k},
 \mathbf{j}\right)\left[ \int_{t'}^{t}ds\mathcal{R}
 \left(k;s,t'\right) \times \right. \nonumber \\
 & & \left. \frac{}{} \times \mathcal{R}\left(j;t,s\right)\mathcal{R}
 \left(\left|\mathbf{k}-\mathbf{j}\right|;t,s\right)C
 \left(\left|\mathbf{k}-\mathbf{j}\right|,s\right)\right] \nonumber \\
 &  & -\theta\left(t-t'\right)\int d^{3}jL\left(\mathbf{k},\mathbf{j}\right)
 \int_{0}^{t'}ds\left\{ \frac{}{}\mathcal{R}\left(k;t',s\right)\mathcal{R}
 \left(j;t,s\right)\times\right.\nonumber \\
 &  & \!\!\!\!\!\!\!
  \left.\times\mathcal{R}\left(\left|\mathbf{k}-\mathbf{j}\right|;t,s\right)
 \frac{C\left(\left|\mathbf{k}-\mathbf{j}\right|,s\right)}
 {C\left(k,t'\right)}\left[C\left(j,s\right)-C\left(k,s\right)\right]\right\} ,
 \label{eq:eta01}
 \end{eqnarray}
 is the turbulent eddy-decay rate and is obtained by comparison with
(\ref{eq:response04}).

Rearranging (\ref{eq:PropLang}) we obtain
\bea
& &  \!\!\!\!\!\!\!
\theta(t-t')\frac{\partial}{\partial t}\mathcal{R}\left(k;t,t'\right) \nonumber \\
& & =-\theta(t-t')\left[\nu
k^{2}+\eta(k;t,t')\right]\mathcal{R}\left(k;t,t'\right), 
\label{eq:PropLang2}
\eea
and this allows us to write (\ref{eq:memtdyneqn01})
as:
\begin{eqnarray}
& & \frac{\partial}{\partial t}D(k,j,\left|\mathbf{k}-\mathbf{j}\right|;t) \nonumber \\ 
& & =1-\int_{0}^{t}ds\left[\frac{}{}\mathcal{R}\left(k;t,s\right)\mathcal{R}
\left(j;t,s\right)\mathcal{R}\left(\left|\mathbf{k}-\mathbf{j}\right|;t,s\right)
\right.\times\nonumber \\
 &  & \times\left[\left(\nu k^{2}+\nu j^{2}+\nu\left|\mathbf{k}-\mathbf{j}\right|^{2}\right)+\eta(k;t,s)\right.\nonumber \\
 &  & \left.\left.\frac{}{}+\eta(j;t,s)+\eta\left(\left|\mathbf{k}-\mathbf{j}
 \right|;t,s\right)\right]\right].
 \label{eq:memtdyneqn02}
 \end{eqnarray}
To be able to calculate (\ref{eq:memtdyneqn02}) we need to take
the Markovian step
\beq
\eta(k;t,s)\rightarrow\eta(k,t).
\eeq
We can justify this step by looking at equations (\ref{eq:Corlink}), (\ref{eq:Mark1tcor})
and (\ref{eq:PropLang}). Equation (\ref{eq:PropLang}) has a general solution
\beq
\mathcal{R}\left(k;t,t'\right)=\exp\left\{ -\nu k^{2}(t-t')-\int_{t'}^{t}ds\eta(k;s,t')\right\} .
\label{eq:propRep03}
\eeq
If we write (\ref{eq:Mark1tcor}) in the suggestive form
\beq
\left[\frac{\partial}{\partial t}+2\nu k^{2}+2\xi(k,t)\right]C\left(k,t\right)=0,
\label{eq:1tCorDyn01}
\eeq
where
\bea
& & \!\!\!\!\!\!\!\!\!
\xi(k,t)=-\int d^{3}jL\left(\mathbf{k},\mathbf{j}\right)
D(k,j,\left|\mathbf{k}-\mathbf{j}\right|;t) \times \nonumber \\ 
& & \times \frac{C\left(\left|\mathbf{k}-\mathbf{j}\right|,t\right)}
{C\left(k,t\right)}\left[C\left(j,t\right)-C\left(k,t\right)\right],
\label{eq:zeta01}
\eea
then we can write the general solution of (\ref{eq:1tCorDyn01}) as
\begin{eqnarray}
& & \!\!\!\!\!\!\!\!\!\!\!\!\!\!\!\!\!\!
C\left(k,t\right) \nonumber \\
& & \!\!\!\!\!\!\!\!\!\!\!\!
= \exp\left\{ -2\nu k^{2}(t-t')-2\int_{t'}^{t}ds\xi(k,s)\right\} 
C\left(k,t'\right)\nonumber \\
 & & \!\!\!\!\!\!\!\!\!\!\!\!
 = \left[\exp\left\{ -\nu k^{2}(t-t')-\int_{t'}^{t}ds\xi(k,s)\right\} \right]^{2}C\left(k,t'\right).
 \label{eq:linkEqn02}
 \end{eqnarray}

Writing (\ref{eq:Corlink}) as
\beq
C(k,t)=\theta(t-t')\mathcal{R}(k;t,t')\mathcal{R}(k;t,t')C(k,t'),
\eeq
and comparing with (\ref{eq:linkEqn02}), this suggests
\beq
\mathcal{R}\left(k;t,t'\right)=\exp\left\{ -\nu k^{2}(t-t')-\int_{t'}^{t}ds\xi(k,s)\right\} .
\label{eq:propRep04}
\eeq

But comparing this with (\ref{eq:propRep03}), we see that
\beq
\int_{t'}^{t}ds\xi(k,s)=\int_{t'}^{t}ds\eta(k;s,t').
\label{eq:ZetaEta}
\eeq
Comparing the forms of $\xi(k,s)$ and $\eta(k;t,s)$, (\ref{eq:zeta01})
and (\ref{eq:eta01}), we see that
\beq
\xi(k,s)=\eta(k;s,s).
\eeq
Also in (\ref{eq:ZetaEta}), since both $t$ and $t'$ are arbitrary,
such that we can make $t\sim t'$, we may make the important assumption
that
\begin{eqnarray}
\eta(k;s,t') & = & \xi(k,s)\nonumber \\
 & = & \eta(k;s,s)=\eta(k,s).
 \end{eqnarray}
This tells us that in the case of the $\eta(k;s,t')$ term, we need
only concern ourselves with the on-diagonal terms
\footnote{Leslie (1973) in deriving an equation for $\eta(k,t)$ from
DIA, averages over the second time argument i.e. $\eta(k,t)=\int_{0}^{t}ds\eta(k;t,s)$,
whereas we simply take the on-diagonal terms. In effect Leslie's $\eta(k,t)$
should be written as $\overline{\eta}(k,t)$ showing that it is an
averaged quatity.}
 $\eta(k;s,s)=\eta(k,s)$, which is a Markovian simplification.

Going back to (\ref{eq:memtdyneqn02}), we can now write it as
\begin{eqnarray}
& & \frac{\partial}{\partial t}D(k,j,\left|\mathbf{k}-\mathbf{j}\right|;t) \nonumber \\
& & = 1-\left[\left(\nu k^{2}+\nu j^{2}+\nu\left|\mathbf{k}-\mathbf{j}\right|^{2}\right)+\eta(k,t)+\eta(j,t)\right.\nonumber \\
 &  & \left.\frac{}{}+\eta\left(\left|\mathbf{k}-\mathbf{j}\right|,t\right)\right]
 D(k,j,\left|\mathbf{k}-\mathbf{j}\right|;t),
 \end{eqnarray}
which along with (\ref{eq:Mark1tcor}) can be used to evolve the memory
time.

\subsection{Single-time Markovianized LET equations}

The final equations for the single-time evolution may now be summarised as:
\begin{eqnarray}
& & \!\!\!\!\!\!\!\!\!
\left(\frac{\partial}{\partial t}+2\nu k^{2}\right)C\left(k,t\right) = 2\int d^{3}jL\left(\mathbf{k},\mathbf{j}\right) \times \nonumber \\
& & \!\!\!\!\!\!\!\!\!
\times D(k,j,\left|\mathbf{k}-\mathbf{j}\right|;t)
C\left(\left|\mathbf{k}-\mathbf{j}\right|,t\right)\left[C\left(j,t\right)-C\left(k,t
\right)\right]\nonumber \\
 & & = -2\eta(k,t)C\left(k,t\right),
 \end{eqnarray}
\bea
\eta(k,t) & = & -\int d^{3}jL\left(\mathbf{k},\mathbf{j}\right)
D(k,j,\left|\mathbf{k}-\mathbf{j}\right|;t) \times \nonumber \\
& & \times \frac{C\left(\left|\mathbf{k}-\mathbf{j}\right|,t\right)}
{C\left(k,t\right)}\left[C\left(j,t\right)-C\left(k,t\right)\right],
\label{eq:EddyDecayer}
\eea
and
\begin{eqnarray}
& & \frac{\partial}{\partial t}D(k,j,\left|\mathbf{k}-\mathbf{j}\right|;t) \nonumber \\
& & = 1-\left[\left(\nu k^{2}+\nu j^{2}+\nu\left|\mathbf{k}-\mathbf{j}\right|^{2}\right)+\eta(k,t)+\eta(j,t)\right.\nonumber \\
 &  & \left.\frac{}{}+\eta\left(\left|\mathbf{k}-\mathbf{j}\right|,t\right)
 \right]D(k,j,\left|\mathbf{k}-\mathbf{j}\right|;t).
 \end{eqnarray}
These equations can be solved numerically with some suitable choice of initial
conditions: 
\begin{eqnarray}
C(k,t=0) & = & \frac{E(k,t=0)}{4\pi k^{2}},
\end{eqnarray}
where $E(k,t=0)$ is an arbitrarily chosen initial energy spectrum, and
\beq
D(k,j,\left|\mathbf{k}-\mathbf{j}\right|;t=0)=0.
\eeq
The last of these initial conditions follows from the definition of
$D(k,j,\left|\mathbf{k}-\mathbf{j}\right|;t)$ in equation (87)
and this in turn implies, from (\ref{eq:EddyDecayer}), that $\eta(k,t=0)=0$,
as is expected, because the cascade has not yet begun at $t=0$.

This set of equations is almost identical to those of the Test Field
Model (TFM) \cite{Kraichnan71b}, the exception being an extra term on
the right hand side of (105) when compared with the corresponding TFM
equation. As before, this extra term guarantees compatibility with
K41. However, it remains to be seen how well the single-time LET theory
performs when computed for the standard test problems.

\section{Conclusion}

We have seen that our time-ordering procedures, as reported in
\cite{Kiyani04} have allowed us to tidy up some aspects of the LET
theory. In particular, we have been able to derive a single-time form of
the theory, which we have Markovianized so that it can be compared with
well-known models such as TFM or EDQNM. Such a comparison will require
numerical computation and this will be the subject of further work.
However, we shall conclude here with some remarks about the role of the
fluctuation-dissipation relation in the Eulrean two-time closures,
DIA, SCF and LET theories.

As we have seen, LET uses the FDR either to derive the response equation
or to be used instead of a response equation. That is, with the second-order
covariance equations for $C(k;t,t')$ and $C(k;t,t)$ we can specify
$R(k;t,t')$ through the FDR and this gives us the requisite set of three
equations.

In contrast, SCF works with (\ref{diag}) for $C(k;t,t)$, the DIA
response equation (\ref{rdia}) for $R(k;t,t')$ and the FDR to calculate
$C(k;t,t')$. Calculations based on these three equations are known to
agree quite closely with those for DIA, consisting of equations
(\ref{dia}), (\ref{diag}) and (\ref{rdia}).

In the case of the DIA, one may test the idea of a FDR by introducing a
modified response function $R'(k;t,t')$, such that
\beq
R'(k;t,t') = \frac{Q(k;t,t')}{Q(k;t',t')}.
\eeq
This quantity plays no part in the calculation. However, at each stage,
$\tilde{R}$ can be calculated from the above relationship and compared
with the actual DIA response function $R$ at the same stage. It is this
comparison that is the basis of the observation that the FDR is quite a
good approximation at smaller wavenumbers but is less good in the
dissipation range \cite{Herring72}. However, such a comparison assumes
that the DIA response equation is `right' and the FDR is `wrong'. In
fact we know that the DIA response equation does not possess the correct
behaviour at large Reynolds numbers and therefore cannot be a standard
of comparison. In our view, the comparison of DIA with LET is a fairer
test of the use of the FDR for turbulence.

Lastly, our derivation of the FDR \cite{Kiyani04} is correct to
second-order in renormalized perturbation theory. Accordingly it is an
approximation, but no more an approximation than the second-order
covariance equations (\ref{dia}) and (\ref{diag}). Therefore its use
with these equations, as in the LET theory, is entirely consistent.
Nevertheless, we should draw a distinction between this situation and
that in microscopic equilibrium systems, where the linear form
(\ref{fdr}) holds to all orders in perturbation theory.

\end{document}